# Where was the particle?

Sofia Wechsler [1)]

**Abstract**

The hypothesis of empty/full waves considers that a click in a detector is triggered by a property carried by the wave-packet that impinges on that detector. Different authors call this property *particle*, but according to the terminology of this hypothesis, the term *full wave* is used in this text.

The present article discusses the validity of the empty/full waves hypothesis. It is shown that the energy conservation principle imposes for a full wave a continuous trajectory from the source to the detector. That gives legitimacy to the question – in case the wave function consists in a couple of space-separated branches – which wave-packet was a full wave before the click in the detector, or, in other authors' terminology, *where was the particle* before the detector clicked. This question is a counterfactual question.

In some multi-particle experiments, e.g. Hardy's thought-experiment with an electron and a positron, the combination of the empty/full waves hypothesis and counterfactual reasoning leads to a contradiction with the predictions of the quantum theory. However, it is shown here that this contradiction is solvable if the locality requirement is relaxed.

To the difference from the local hidden variables, it is not possible to prove or disprove the idea of empty/full waves, at least at present. But it is possible to ask whether this hypothesis offers a solution to the quantum puzzles. The answer is negative. For instance it offers no solution to the before-before loop.

The question remains open whether the empty/full waves hypothesis is correct, or "God plays dice".

## 1. Introduction

According to the hypothesis of empty/full waves, a click in a detector is triggered by a property carried by the wave-packet that impinges on that detector. This hypothesis stems from our hope that in the detection process, a recording in a detector is triggered by some property of the impinging wave-packet, not by properties of the detector and/or of the environment. Otherwise the result is undefined; it is bound to differ from detector to detector or to change with changes in the environment.

The empty/full waves idea was studied by different authors that predicted effects in contradiction with the quantum theory [1]. The experiment disproved such effects placing the empty/full waves idea under question mark [2]. However, L. Hardy showed that if the empty waves exist, they are bound to have an observable effect that doesn't contradict the quantum theory, [3]. What he proved was a uni-directional implication: *if* empty waves exist they have an observable effect; however, that effect does not require with necessity the empty waves hypothesis – the explanation of the effect may be different. Besides that, in another article that became famous [4] Hardy described a thought-experiment that seems to disprove the idea of empty waves.[2)] The conservation energy implies that a full wave should follow a continuous trajectory from the source to the detector. This conclusion justifies questions as *where was* the full wave before the click in the detector, in case that more than one trajectory are possible. This is a counterfactual question. In the end of such a counterfactual rationale, Hardy arrived at a contradictory situation, known in literature as "Hardy's paradox". Berndl and Goldstein [5] claimed that the contradictory situation is due to the counterfactual reasoning. However, as shown below, there are different ways to reason counterfactually, not necessarily the one chosen by Hardy, s.t. the contradiction may be solved. But the contradiction may also be solved if one

---

[1)] Computers Engineering Center, Nahariya, P.O.B. 2004, 22265, Israel
[2)] Hardy spoke of "particles", not of empty/full waves, considering the wave function as a non-local guide-wave for the particle, in the spirit of de Broglie's theory. In this text we speak of full waves.

abandons the hypothesis of empty/full waves, and with that, the justification for counterfactual reasoning. The bottom line, it is not yet clear if this hypothesis is correct or not. In addition, another challenge stands in front of this hypothesis. In two-particle entanglements the quantum correlations indicate that the measurement outcome of one particle depends on the type and result of the test done on the other particle. But the result of the other particle also depends on the type and result of the test on its pair-particle. This is a closed loop, with no visible way out, than assuming asymmetry in the behavior of the two particles, i.e. that one of them should behave independently, and the other one dependently [6], [7]. It is shown in the last section that the empty/full waves hypothesis gets stuck with this loop.

The rest of the text is organized follows. Section **2** proposes an implementation of Hardy's experiment without applying to the electron-positron annihilation reaction which has a too small probability to occur. The proposed implementation is inspired by indications given by Hardy in another article, [8]. Section **3** describes a clash with the quantum theory. Section **4** shows that the clash may be solved without rejecting the empty/full waves, but for the above closed loop this hypothesis proves to be useless.

## 2. An implementation of Hardy's experiment without annihilation

Consider a pair of down-conversion photons, named here $e^+$ and $e^-$ to preserve the notations in [4]. The pair is prepared in the initial state

(1) $|\Psi> = 2^{-\frac{1}{2}}(\ |a^+>|a^-> + |b^+>|b^->)$

where $\mathbf{a}^+$, $\mathbf{b}^+$, are the paths followed by the wave-packets of the photon $e^+$, and $\mathbf{a}^-$, $\mathbf{b}^-$, the paths followed by the wave-packets of the photon $e^-$. The beams $\mathbf{a}^+$, $\mathbf{a}^-$, $\mathbf{b}^+$, $\mathbf{b}^-$, impinge on the four beam-splitters $\mathbf{BS1}^{\pm}$ as shown in fig. 1a. These beam-splitters transmit ⅓ from the incident intensity and the rest is reflected. To the difference, the beam-splitters $\mathbf{BS}^{\pm}$ transmit and reflect equally. Taking also in consideration the phase shifts of $^{3\pi}/_2$ on the rays reflected by the mirrors $\mathbf{M}^{\pm}$, one has

(2) $|a^{\pm}> \rightarrow 3^{-\frac{1}{2}}(\ |v^{\pm}> + i|u^{\pm}> - |g^{\pm}>)$ , $\quad |b^{\pm}> \rightarrow 3^{-\frac{1}{2}}(|f^{\pm}> - |u^{\pm}> + i|g^{\pm}>)$ ,

Introducing (2) in (1) one gets

(3) $|\Psi> = 6^{-\frac{1}{2}}\{3^{-\frac{1}{2}}(\ |v^+>|v^-> + i|v^+>|u^-> + i|u^+>|v^->) + A(|g^{\pm}>, |f^{\pm}>)\}$,

where $A(|g^{\pm}>, |f^{\pm}>)$ is an expression in which every term comprises at least one of the factors $|g^+>$, or $|g^->$, or $|f^+>$, or $|f^->$, therefore these cases may be easily identified and discarded. The expression to be studied is

(4) $|\psi> = 3^{-\frac{1}{2}}(\ |v^+>|v^-> + i|v^+>|u^-> + i|u^+>|v^->)$ .

That completes the preparation step of the experiment.

In continuation, the rays $\mathbf{u}^{\pm}$, $\mathbf{v}^{\pm}$, impinge on the fair beam splitters $\mathbf{BS2}^{\pm}$, fig 1b. They produce the transformations

(5) $|u^{\pm}> \rightarrow 2^{-\frac{1}{2}}(\ |c^{\pm}> + i|d^{\pm}>)$ , $\quad |v^{\pm}> \rightarrow 2^{-\frac{1}{2}}(\ |d^{\pm}> + i|c^{\pm}>)$ .

The detectors in fig. 1b are supposed ideal and mobile. They may be displaced at the free will of the experimenter in the respective region. If the detectors in the region '+' are displaced to the paths $\mathbf{u}^+$, $\mathbf{v}^+$, the wave function that describes the possible outcomes of the measurement is obtained by substituting in (4) the transformations (5) for $e^+$

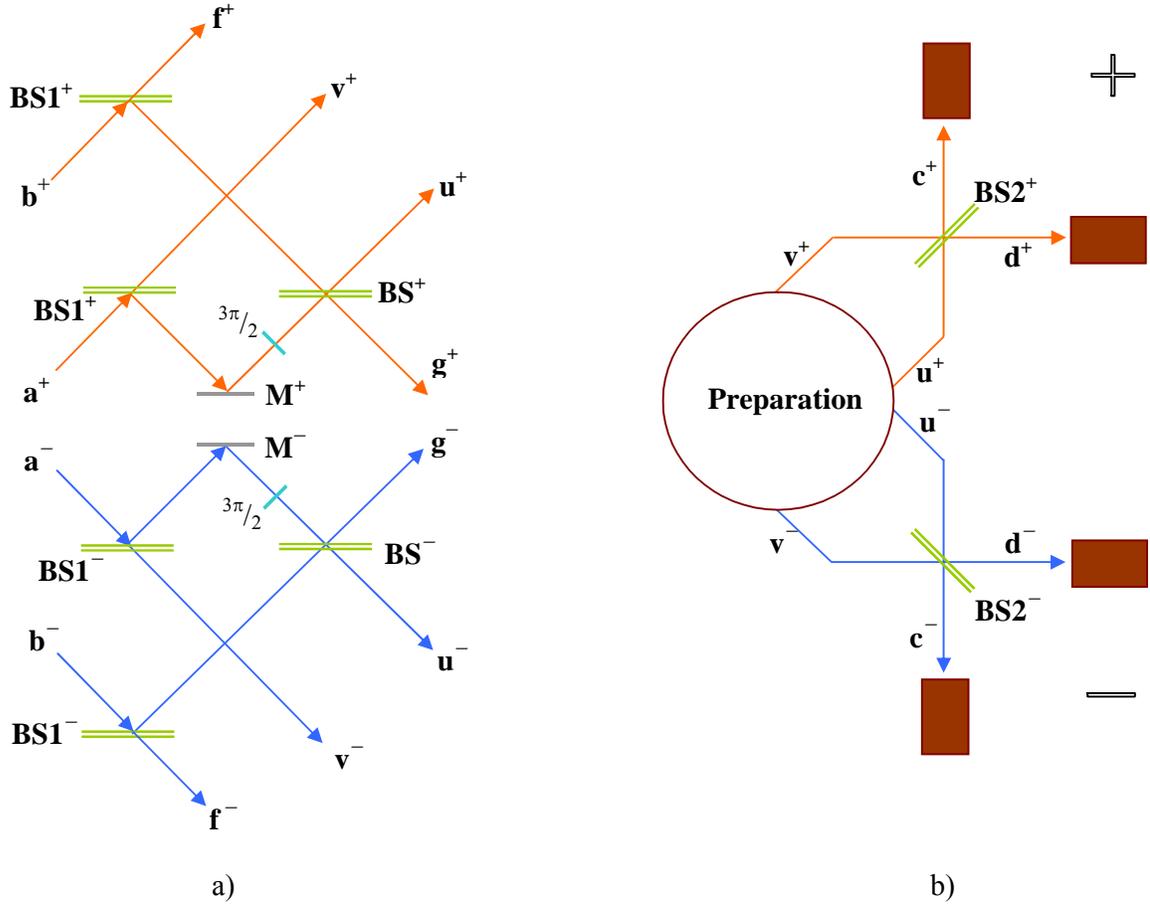

**Figure 1. An implementation of Hardy's experiment.**
The two photons are marked with different colors for eye-guiding.
a) Preparation; b) Detection.

(6) $|\psi\rangle = 6^{-\frac{1}{2}}( 2i|c^+\rangle|v^-\rangle + i|d^+\rangle|u^-\rangle - |c^+\rangle|u^-\rangle )$,

Similarly, if the detectors in the region '–' are displaced to the paths $u^-$, $v^-$, the wave function that describes the measurement outcomes is obtained by substituting in (4) the transformations (5) only for $e^-$

(7) $|\psi\rangle = 6^{-\frac{1}{2}}( 2i|v^+\rangle|c^-\rangle + i|u^+\rangle|d^-\rangle - |u^+\rangle|c^-\rangle )$.

Finally, if all the detectors are in their places as in fig. 2b,

(8) $|\psi\rangle = 12^{-\frac{1}{2}} ( -3|c^+\rangle|c^-\rangle + i|c^+\rangle|d^-\rangle + i|d^+\rangle|c^-\rangle - |d^+\rangle|d^-\rangle )$.

## 3. A difficulty

Before proceeding to further analysis, some immediate remarks on the full/empty waves are possible:

(a) *At any moment, one and only one of the two wave-packets of a photon is a full wave.*
(b) *A full wave cannot jump from one track to another.*

The justification of the remark (a) is that otherwise, one single photon will be detected in two detectors, or in no detector – each variant contradicting the energy conservation. The justification of the remark (b) is that otherwise, a moving frame of coordinates may be found in which the jump is done with finite velocity, s.t. during the jump interval the photon won't be detectable on any track; that again contradicts the energy conservation.

In consequence of the wave function (4) there are three possibilities for the full wave-packets at the exit of the preparation region: 1) to travel along the pair of paths $\mathbf{u}^+$, $\mathbf{v}^-$; 2) along $\mathbf{v}^+$, $\mathbf{v}^-$; 3) along $\mathbf{v}^+$, $\mathbf{u}^-$.
With the scenarios 1) and 2) a problem arises assuming non-local influence. One can see from (6) that if the full wave of $e^-$ goes to $\mathbf{v}^-$, then later on, when leaving the beam-splitter $\mathbf{BS2}^+$, the full wave of $e^+$ will have to go to $\mathbf{c}^+$. In consequence one couldn't get in these two cases a detection on $\mathbf{d}^+$.
The scenario 3) may permit a detection on $\mathbf{d}^+$, but eq. (7) shows that if the full wave of $e^+$ goes to $\mathbf{v}^+$, then later on, when leaving the beam-splitter $\mathbf{BS2}^-$, the full wave of $e^-$ will be guided by an analogous non-local influence, to $\mathbf{c}^-$. Therefore a detection on $\mathbf{d}^-$ is not possible.
In all, no scenario leads to a joint detection on $\mathbf{d}^+$ and $\mathbf{d}^-$. But the experiment was performed by Irvine et al. [9] and cases of the type $\mathbf{d}^+$, $\mathbf{d}^-$, were found according to the predictions of the wave function (8).

## 4. Discussion

Berndl and Goldstein [5] considered that the above difficulty stems from the counterfactual reasoning which is in disagreement with the *contextuality principle*. This principle is a consequence of Gleason's theorem [10], and was first formulated by Bell [11], then developed by Kochen and Specker [12] and Peres [13]. According to this principle, in a Hilbert space of dimension ≥ 3 the result of a measurement depends on the type of the other measurements done on the system studied. In our case we have a vector space of dimension 4. While the detectors in the region '+' stand on the paths $\mathbf{c}^+$ and $\mathbf{d}^+$, the detectors in the region '–' stand on the paths $\mathbf{c}^-$ and $\mathbf{d}^-$, not on $\mathbf{u}^-$ and $\mathbf{v}^-$, and in consequence, the wave function (6) should not be used to predict results, fig. 2. Analogously, while the detectors in the region '–' stand on $\mathbf{c}^-$ and $\mathbf{d}^-$, the detectors in the region '+' stand on $\mathbf{c}^+$ and $\mathbf{d}^+$, not on $\mathbf{u}^+$ and $\mathbf{v}^+$, therefore, the wave function (7) should not be used to predict results.
Thus, as long as our detectors are on $\mathbf{c}^+$, $\mathbf{d}^+$, $\mathbf{c}^-$, $\mathbf{d}^-$, the contextuality principle releases us of the constraints implied by the wave functions (6) and (7). The fact may be justified if the locality is abandoned, since by the time a detection takes place on one of the paths $\mathbf{c}^+$, $\mathbf{d}^+$, at least according to the time axis of the lab, the wave packets of the photon $e^-$ exited $\mathbf{BS2}^-$; so, the wave function is (8), and it encapsulates the information that no detector was on $\mathbf{u}^-$ and $\mathbf{v}^-$.
The empty/full waves hypothesis and the idea of trajectories are not in conflict with the contextuality.

But, borrowing an expression used once by A. Shimony, "*Triumph is short lived*".
Let's return to fig. 1a and remove all the beam-splitters, mirrors, and phase shifters, and input the beams $\mathbf{a}^\pm$ into $\mathbf{u}^\pm$, and the beams $\mathbf{b}^\pm$ into $\mathbf{v}^\pm$. The expression (1) becomes $2^{-\frac{1}{2}}(\ |\mathbf{u}^+>|\mathbf{u}^->+|\mathbf{v}^+>|\mathbf{v}^->\ )$. Then, using the transformations (5) the state beyond $\mathbf{BS2}^\pm$ should be $2^{-\frac{1}{2}}i(\ |\mathbf{c}^+>|\mathbf{d}^->+|\mathbf{d}^+>|\mathbf{c}^->\ )$, i.e. the full waves should go either to $\mathbf{c}^+$ and $\mathbf{d}^-$, or to $\mathbf{d}^+$ and $\mathbf{c}^-$. The question is how that is done.
Assume that the full waves travel in the beginning on $\mathbf{u}^+$, $\mathbf{u}^-$. At the exit of the beam-splitters $\mathbf{BS2}^\pm$ they can't pick freely one path out of $\mathbf{BS2}^+$ and one path out of $\mathbf{BS2}^-$, because the results would be uncorrelated. The path taken by the full wave of $e^+$ depends on the path taken by the full wave of $e^-$, and vice-versa. This is a dead-lock and the two full waves should stop moving. But this is not what happens, and the empty/full waves hypothesis is unable to explain why.
This problem was examined by different researchers. A. Suarez and his colleagues [6] proposed the idea that the equivalence in the two particles behavior is broken: one of them produces its response independently. For instance he suggested that if the particle $e^+$ crosses the beam-splitter $\mathbf{BS2}^+$ before the particle $e^-$ reaches the beam-splitter $\mathbf{BS2}^-$, then $e^+$ produces its response independently. Analogously for $e^-$. The meaning of the word "before" was proposed to be given by the time axis in the rest-frame of the respective beam-splitter. The experiment was performed with moving beam-splitters. The responses of the particles were expected to be non-correlated because with the above time-convention, each particle was the first to cross the beam-splitter – hence the name before-before for this experiment. However, the correlations appeared as predicted by the quantum theory.

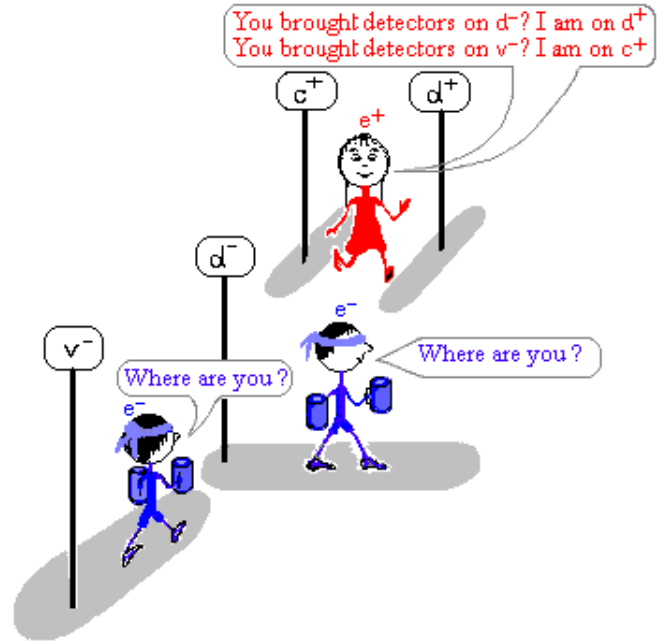

**Figure 2. Contextuality.**

It may be argued that other strategies of breaking the equivalence between the particles may be chosen,[3] or other choices may be suggested for the time axes – some people support the idea of preferred frames of coordinates. The before-before experiment didn't rule out the empty/full wave hypothesis, but proved it as useless vis-à-vis this problem.

---

[3] N. Gisin performed an experiment not based on empty-waves [7]. Admitting that the measurement results are established at the contact of the wave-packets with the classical detectors, and not beforehand, he supposed that the particle first to meet a classical detector might produce its response independently. Simulating part of the detectors by absorbers, each particle was the first to meet an absorber according to the rest-frame of that absorber. With one of the absorbers in movement the correlations should have been violated. But the correlations persisted.